\documentclass[a4paper,11pt]{article}
\pdfoutput=1 
\usepackage{jheppub} 

\usepackage[T1]{fontenc} 
\usepackage[toc,page]{appendix}
\usepackage{empheq}
\usepackage{graphicx}
\usepackage{subcaption}
\usepackage{todonotes}

\newcommand{\cN}{\mathcal{N}}

\newcommand{\vp}{\varphi}

\def\K{K{\"a}hler}
\def\be{\begin{equation}}
\def\ee{\end{equation}}
\newcommand{\ba}{\begin{eqnarray}}
\newcommand{\ea}{\end{eqnarray}}

\newcommand{\rf}[1]{(\ref{#1})}

 \title{\huge {\rm {\bf  {\boldmath {Hypernatural inflation}}}}}

\author[a]{Andrei Linde,}
\author[b,c]{Dong-Gang Wang,}
\author[b,c]{Yvette Welling,}
\author[a]{Yusuke Yamada}
\author[b,d]{and Ana Ach\'ucarro}

\affiliation[a]{Stanford Institute for Theoretical Physics and Department of Physics, Stanford University, Stanford, CA 94305, USA}
\affiliation[b]{Lorentz Institute for Theoretical Physics, Leiden University, 2333CA Leiden, The Netherlands}
\affiliation[c]{Leiden Observatory, Leiden University, 2300 RA Leiden, The Netherlands}
\affiliation[d]{Department of Theoretical Physics, University of the Basque Country, 48080 Bilbao, Spain}

\emailAdd{alinde@stanford.edu}
\emailAdd{wdgang@strw.leidenuniv.nl}
\emailAdd{welling@strw.leidenuniv.nl}
\emailAdd{yusukeyy@stanford.edu}
\emailAdd{achucar@lorentz.leidenuniv.nl}

\notoc

\abstract{We construct a model of natural inflation in the context of $\alpha$-attractor supergravity, 
in which both the dilaton field   and the axion field    are light during inflation, and the inflaton may be a combination of the two. The T-model version of this theory is defined on the Poincar\'e disk with  radius $|Z| = 1$. It describes a Mexican hat potential with the flat axion direction corresponding to a circle of radius $|Z|  < 1$. The  axion decay constant $f_{a}$ in this theory can be exponentially large   because of the hyperbolic geometry of the Poincar\'e disk.  Depending on initial conditions, this model may describe  $\alpha$-attractor inflation driven by the radial component of the inflaton field,  natural inflation driven by the axion field, or a sequence of these two regimes. We also construct the E-model version of this theory, which have similar properties.   In addition, we  describe generalized $\alpha$-attractor models where the potential can be singular at the boundary of the moduli space, and show that they can provide a simple solution for the problem of initial conditions for the models with plateau potentials. }

\begin{document}

\maketitle

\newpage

 \tableofcontents{}

\parskip 8pt

\section{Introduction}

In this paper we will continue our investigation of multifield  $\alpha$-attractors following the earlier work \cite{Achucarro:2017ing}. Cosmological $\alpha$-attractors form a broad recently discovered class of inflationary models \cite{Kallosh:2013hoa,Ferrara:2013rsa,Kallosh:2013yoa,Cecotti:2014ipa,Galante:2014ifa,Kallosh:2016gqp,Kallosh:2015zsa,Ferrara:2016fwe,Kallosh:2017ced,Kallosh:2017wnt}. These models give nearly model-independent cosmological predictions $n_s\approx 1-{2\over N}$, $r\approx   {12\alpha \over N^2}$, providing a very good match to the recent observational data \cite{Ade:2015lrj, Ade:2015ava}.

These models are most naturally formulated in terms of the theory of scalar fields with hyperbolic geometry ~\cite{Kallosh:2015zsa,Ferrara:2016fwe,Kallosh:2017ced}. For example,
in the simplest supergravity embedding of these models, the potential  depends on the complex scalar $Z= |Z|  \, e^{i\theta} $, where $Z$ belongs to the Poincar\'e disk with $|Z|   < 1$,   and the kinetic terms are
\be
 3\alpha {\partial_\mu \bar Z \partial^\mu Z \over (1 - Z \bar Z)^2  } \ .
\label{geometry}
\ee
 In many versions of these models, the field $\theta$ is heavy and stabilized at $\theta = 0$,  so that the inflationary trajectory corresponds to the evolution of the single real field $Z = \bar Z$. This field is not canonically normalized, but one can easily express everything in terms of the canonically normalized field $\varphi$, where
 \be\label{ZZ}
Z = \tanh{\vp\over \sqrt{6\alpha}} \ .
 \ee
Then the potential of the inflaton field $V(Z,\bar Z)$ along the inflaton direction can be represented as
\be\label{TT}
V = V(\tanh{\vp\over \sqrt{6\alpha}}) \ .
\ee
Since $\tanh{\vp\over \sqrt{6\alpha}}\to \pm 1$ in the limit $\vp \to \pm \infty$  (corresponding to $Z\to \pm 1$), inflationary predictions are mainly determined by the behavior of $V(Z,\bar Z)$ at the boundary of the moduli space $|Z| \to 1$. This explains the stability of these predictions with respect to modifications of the original potential $V(Z,\bar Z)$ everywhere outside a small vicinity of the points $Z  = \pm 1$.

However, there is another class of models, where the field $\theta$ is not stabilized and may remain  light during the cosmological evolution.
A particularly interesting case is $\alpha=1/3$, which  has a fundamental origin from maximal $\cN=4$ superconformal symmetry and from  maximal $\cN=8$ supergravity  \cite{Kallosh:2015zsa}.  For $\alpha=1/3$, a class of supergravity embeddings are known to possess an unbroken or slightly broken $U(1)$ symmetry, which makes both $\vp$ and $\theta$ light \cite{Kallosh:2015zsa,Achucarro:2017ing}. In that case, the inflationary evolution may involve both fields, which would require taking into account the multi-field effects, as discussed in  \cite{Gordon:2000hv, GrootNibbelink:2000vx, GrootNibbelink:2001qt, Bartolo:2001rt, Lalak:2007vi, Achucarro:2010jv, Achucarro:2010da, Peterson:2010np}. However, in contrast with the naive expectation, the cosmological predictions of the simplest class of such models are very stable not only with respect to modifications of the potential of the field $|Z|$, but also with respect to strong modifications of the potential of the field $\theta$ \cite{Achucarro:2017ing}. 
The predictions coincide with those of the single-field $\alpha$-attractors: $n_s\approx 1-{2\over N}$, $r\approx   {12\alpha \over N^2}$.
Thus, this class of models exhibits double-attractor behavior, providing universal cosmological predictions, which are stable with respect to strong modifications of the potential  $V(Z,\bar Z)$. Other closely related investigations can be found in \cite{Kallosh:2013daa,Christodoulidis}.

This result is valid  for all models with hyperbolic geometry \rf{geometry} and an unbroken or slightly broken $U(1)$ symmetry, for any $\alpha \lesssim O(1)$  \cite{Achucarro:2017ing}. In supergravity such models were known only for $\alpha  = 1/3$, but  recently such models were constructed for all $\alpha < 1$ \cite{Yamada:2018nsk}, so now we have a broad class of supergravity models with the double attractor behavior found in   \cite{Achucarro:2017ing}.

In this paper, we will further generalize this construction. In the models considered in  \cite{Achucarro:2017ing}, we had only one stage of inflation, which ended at $\vp = 0$.  Instead of that, we will consider Mexican hat potentials $V(\vp, \theta)$ with a pseudo-Goldstone direction along the minimum  with respect to the field $\vp$, see Figs. \ref{fig:Higgs}, \ref{fig:Higgs2}. In such models the first stage of $\alpha$-attractor inflation driven by the   field $\vp$  can be followed by a subsequent stage of inflation driven by the field $\theta$, as in the natural inflation scenario \cite{Freese:1990rb}. This leads to a novel realization of the natural inflation scenario in supergravity. 

 Natural inflation is already in tension with the latest  CMB data, except possibly for super-Planckian values of the axion decay constant, which are difficult (if not impossible) to achieve in string theory \cite{Banks:2003sx, Conlon:2012tz}. A number of extensions of natural inflation have tried to address this problem
in different ways  \cite{Kim:2004rp, Dimopoulos:2005ac, Czerny:2014qqa, Li:2015mwa, Achucarro:2015rfa, Achucarro:2015caa, Conlon:2012tz, Cicoli:2014sva,Bachlechner:2014gfa,Bachlechner:2015qja,Long:2016jvd,Dias:2018pgj}. Here,
the hyperbolic field metric results in an effective axion decay constant that is very different from the naive one. Our construction is embedded in $N=1$ supergravity, but whether it
admits an embedding in string theory remains an open question here as much as for other extensions of natural inflation.   We will not discuss this question in our paper, 
trying to solve one problem at a time.

Indeed,  theoretical problems with natural inflation appear already at the level of supergravity. Historically, one could always postulate the periodic potential required for natural inflation \cite{Freese:1990rb}. But it was very difficult to stabilize the inflaton direction in supergravity-based versions of natural inflation.  This problem was solved only few years ago   \cite{Kallosh:2014vja} (see also \cite{Kallosh:2007ig,Kallosh:2007cc}). However, in all natural inflation models in supergravity constructed in \cite{Kallosh:2014vja}, the inflaton field was different from the angular direction $\theta$ in the Mexican hat potential as originally proposed back  in   \cite{Freese:1990rb}. This problem, which remained unsolved since 1990,  will be solved in our paper.

However, as we will discuss in the end of the paper, embedding of natural inflation in the theory of $\alpha$-attractors   does not remove the tension between natural inflation and the latest observational data. Moreover, adding a stage of natural inflation {\it after} the stage of inflation driven by $\alpha$-attractors tends to decrease the predicted value of $n_{s}$. From this perspective,  inflationary models  where the last 60 e-foldings of inflation are driven by $\alpha$-attractors, without an additional stage of natural inflation, seem to provide a better fit to the existing observational data \cite{Ade:2015lrj, Ade:2015ava}.




\section{\boldmath{U(1)-symmetric potentials for $\alpha$-attractors in supergravity:  T-models in disk variables $Z$}} \label{sT}

There are two different ways to obtain $U(1)$ symmetric potentials for $\alpha$ attractors in supergravity. The simplest one is to consider models with $\alpha = 1/3$ with the \K\ potential and superpotential
\be\label{k1}
K=-  \ln (1-Z\bar Z -S\bar S) \ , \qquad W= S\, f(Z) \ ,
\ee
where   $S$ is a nilpotent field and $f(Z)$ is a real holomorphic function.
In these models, the potential $V(Z,\bar Z) = |f(Z)|^{2}$. For the simplest functions $f(Z) = Z^{n}$, we find $\theta$-independent potentials $V(Z,\bar Z) = |f(Z)|^{2} = f^{2}(\tanh{\vp\over \sqrt{6\alpha}})$.  However, for more complicated functions, such as $1-Z^{2}$, which we may need to reproduce the natural inflation potential, the $U(1)$ symmetry of the potential is strongly broken.  Therefore it is essential for us to use the novel method developed in  \cite{Achucarro:2017ing} for $\alpha = 1/3$ and generalized in  for all $\alpha < 1$ \cite{Yamada:2018nsk}.  For a detailed discussion of this construction we refer the readers to the original papers where this formalism was developed  \cite{Achucarro:2017ing,Yamada:2018nsk}, and also to the closely related works \cite{McDonough:2016der,Kallosh:2017wnt}. Here is a short summary of the results required for our work.

Following  \cite{Yamada:2018nsk}, we will consider supergravity with the \K\ potential and superpotential
\be
K=-3\alpha\log(1-Z\bar Z)+S+\bar S+G_{S\bar S}S\bar S \ , \qquad W=W_0 \ .\label{KW}
\ee
The function $G_{S\bar S}$ is given by
\be
G_{S\bar S}=\frac{|W_0|^2}{(1-Z\bar Z)^{3\alpha} {V}(Z,\bar Z)+3|W_0|^2(1-\alpha Z\bar Z)}.
\ee
In this setting, the inflaton potential is given by $V(Z,\bar Z)$, where $ V(Z,\bar Z)$ is an arbitrary real function of $Z$ and $\bar{Z}$.

This formulation  is valid for any $\alpha < 1$.  For the special case $\alpha = 1/3$,  it coincides with the formulation given in  \cite{Achucarro:2017ing}. A detailed explanation of notations and  basic principles of this general approach to inflation in supergravity can be found in \cite{Kallosh:2017wnt}. In our paper, we will focus on constructing specific potentials $ V(Z,\bar Z)$ suitable for implementation of natural inflation in this context.

The kinetic term, which plays important role in this scenario ,  is given by
\begin{equation}\label{kin}
3\alpha\frac{\partial Z \partial \bar{Z}}{(1-Z\bar{Z})^2} = \frac{1}{2}\left(\partial \vp\right)^2 + \frac{3\alpha}{4} \sinh^2 \left(\sqrt{\frac{2}{3\alpha}}\vp\right)\left(\partial \theta\right)^2 \ .
\end{equation}

Importantly, for all functions $V$ depending only on the product $Z\bar Z$, the potential is $U(1)$ invariant, i.e. it does not depend on the angular variable $\theta$. That is exactly what we need as a first step towards the theory of natural inflation: a potential with a flat Goldstone direction.

\begin{figure}[h!]
\centering
		 \includegraphics[width=0.5\textwidth]{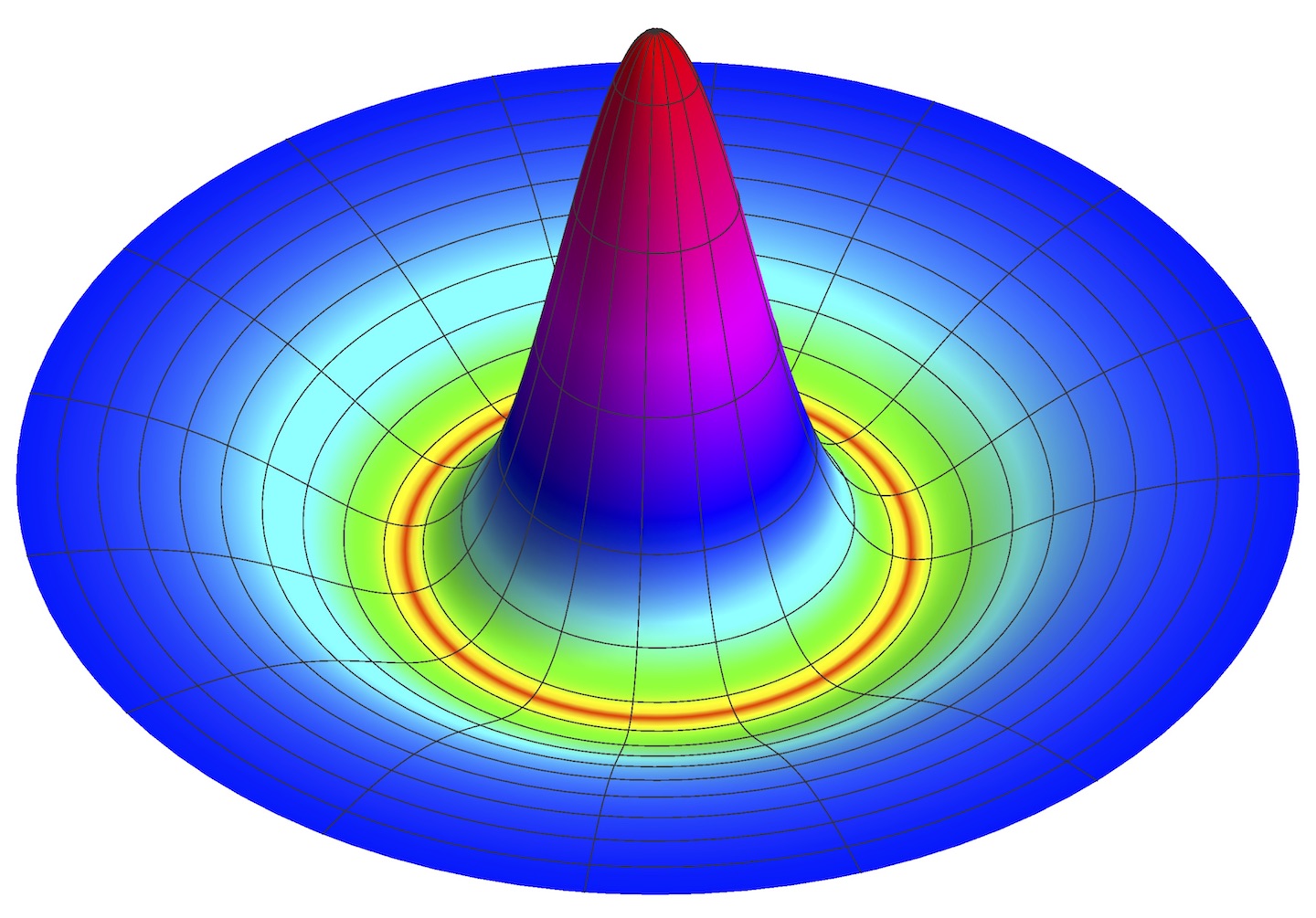}
        \caption{The shape of the hyperbolic generalization of the Higgs potential $V_0  \left(1-c^{-2}\tanh^{2} \frac{\vp}{\sqrt{6\alpha}} \right)^{2}$ for $c = 0.8$ and $\alpha = 1/3$.}
        \label{fig:Higgs}
\end{figure}

To give a particular example, we will begin with the Higgs-type potential
\be\label{h}
V(Z, \bar Z) =V_0 (1- c^{-2}\, Z\bar Z)^{2}
\ee
with $|Z|< 1$.
In terms of $\vp$ and $\theta$, the potential is given by
\be \label{toymodelpotential}
V(\vp, \theta)=V_0 \left(1-c^{-2}\tanh^{2} \frac{\vp}{\sqrt{6\alpha}} \right)^{2}~.
\ee
One may call it ``hyperhiggs'' potential. It has a $\theta$-independent minimum at
\be
\label{minimum}
\tanh\frac{\vp}{\sqrt{6\alpha}}  = c \ ,
\ee
which is shown as a red circle in Fig. \ref{fig:Higgs}.

\section{Natural inflation in disk variables: T-models with a Mexican hat potential}
\label{sec:toy}

As a next step, we will introduce a small term breaking the $U(1)$ symmetry. Here again one can use several different methods. For example, one can add a   term proportional to $ Z$ to the superpotential in \eqref{KW}. One can show that this leads to the natural inflation potential. Instead of that, we will use a simpler phenomenological method, which we already used in   \cite{Achucarro:2017ing}: We will add to the hyperhiggs potential \rf{h} a small term containing a sum $Z^n+\bar{Z}^n$, which breaks the $U(1)$ symmetry.

\begin{figure}[h!]
\centering
		 \includegraphics[width=0.32\textwidth]{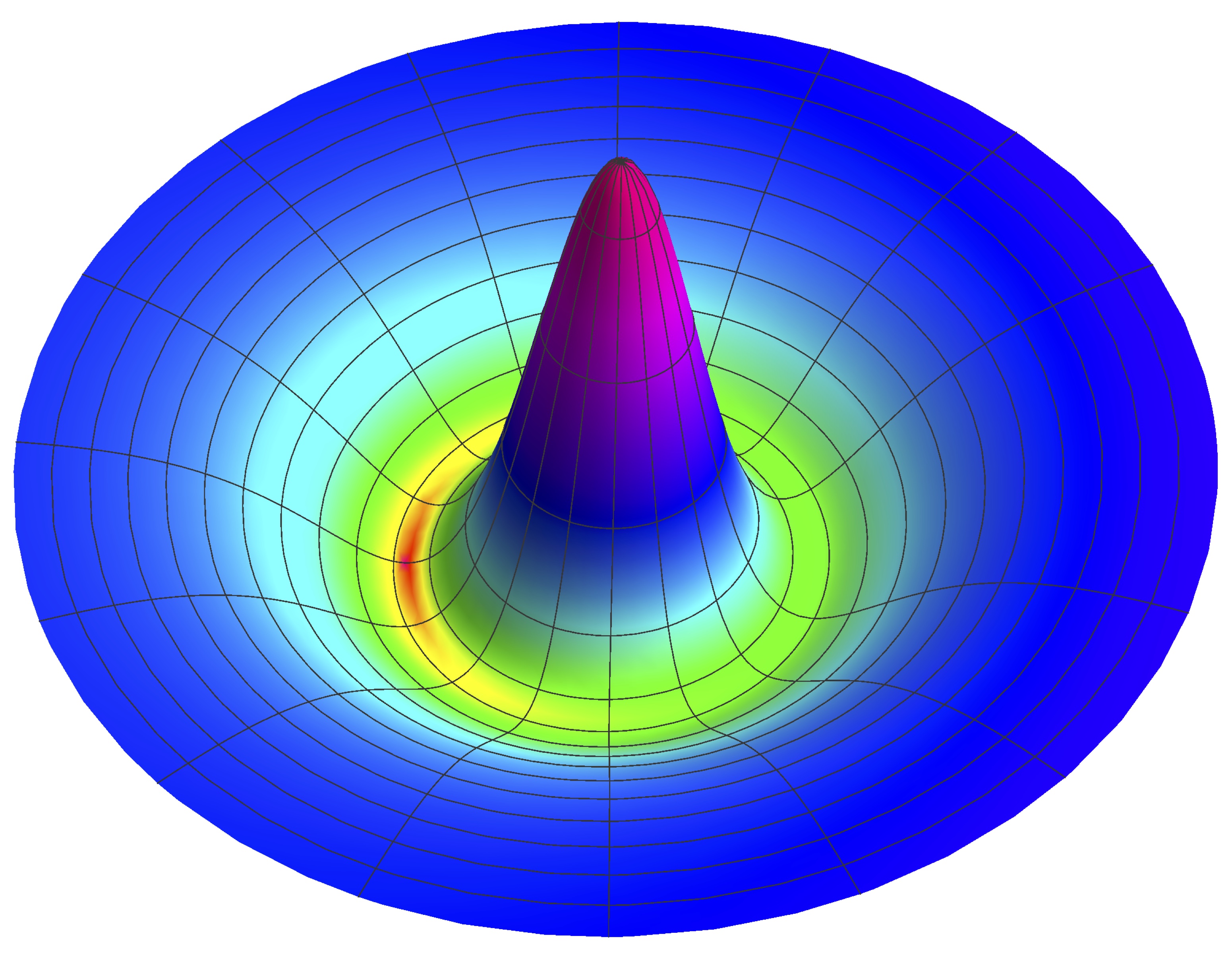}\quad
		 \includegraphics[width=0.32\textwidth]{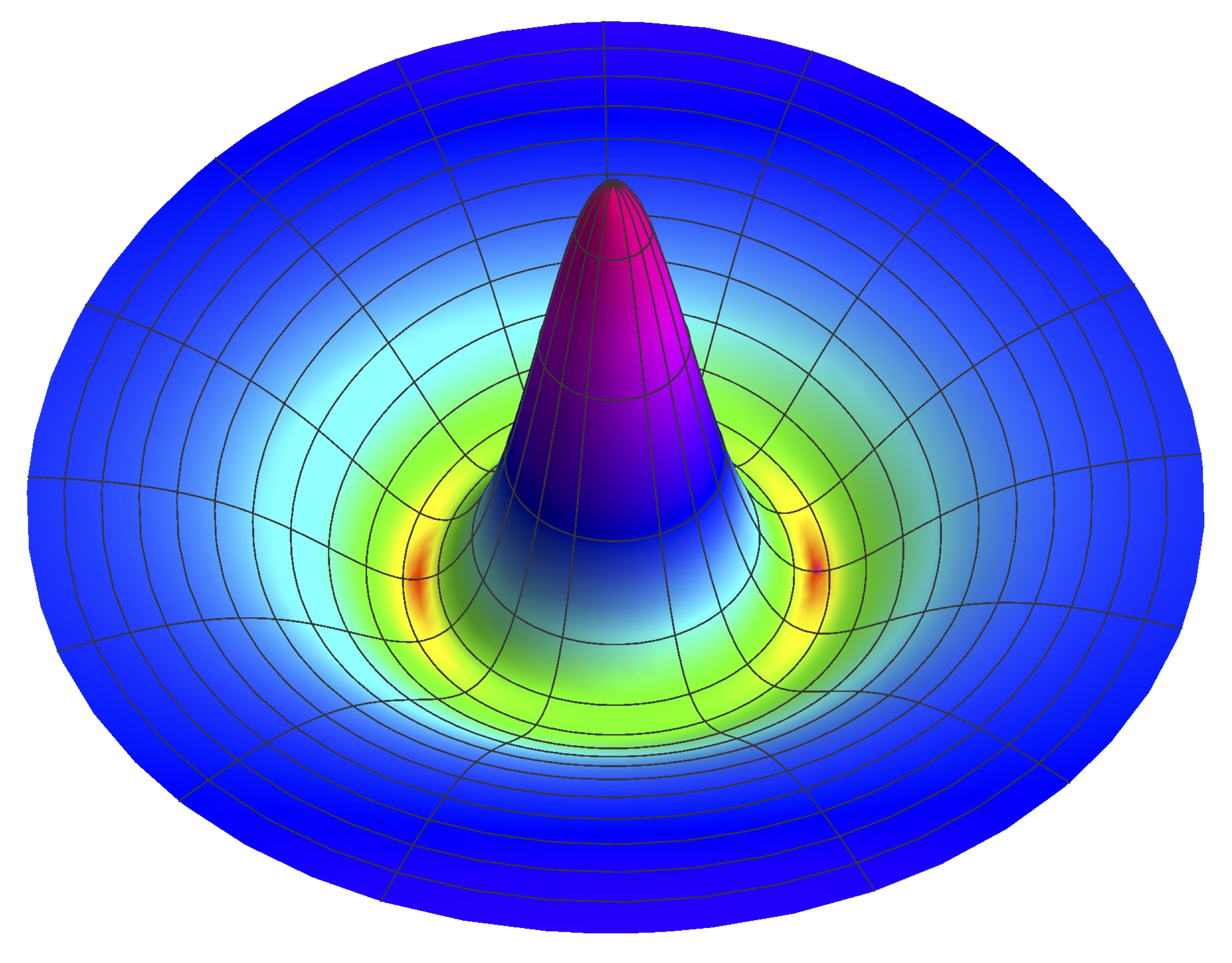}\quad
		 \includegraphics[width=0.30\textwidth]{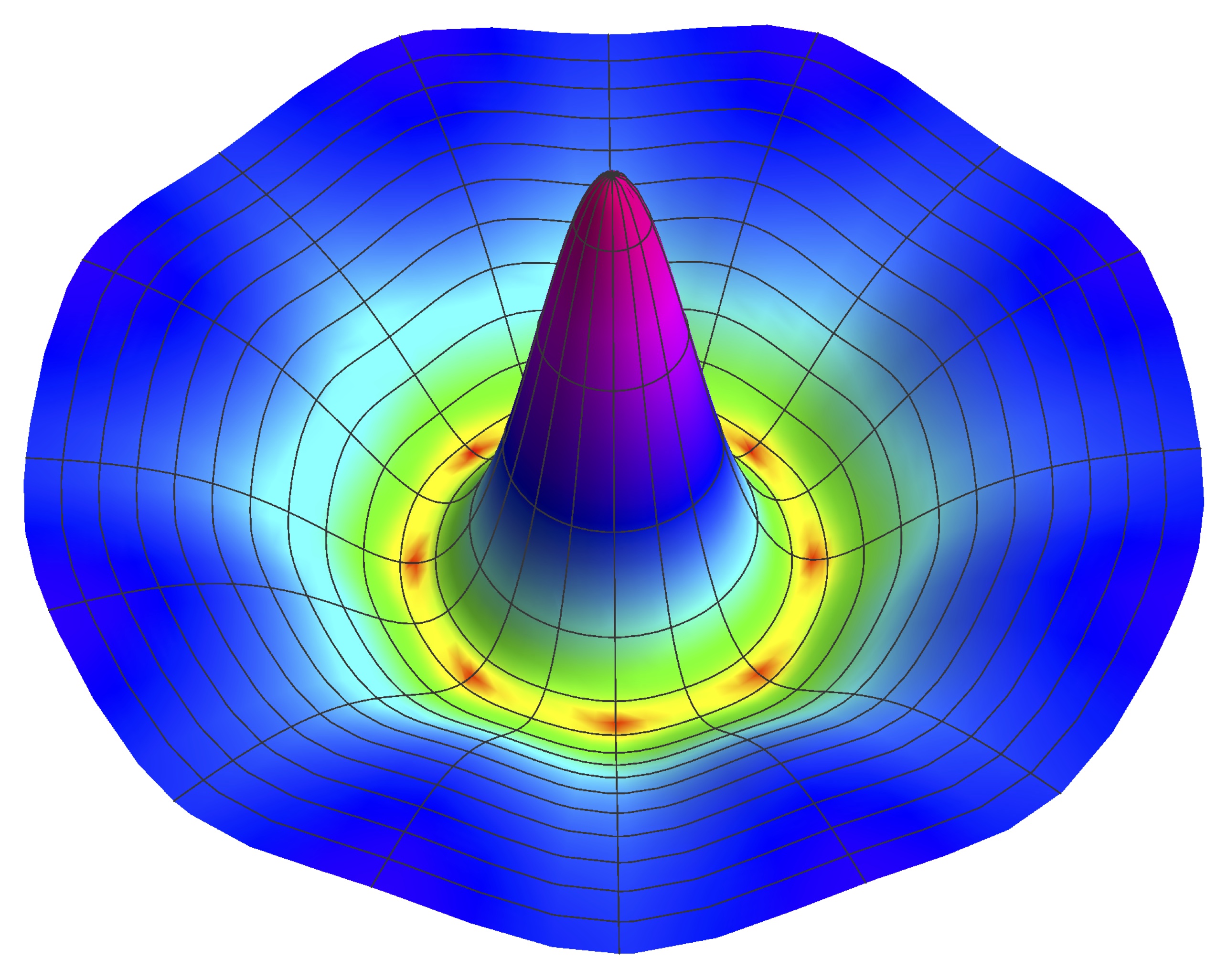}
        \caption{The natural inflation potential \rf{toymodelpotential2} with one, two, and eight minima (shown by red) along the pseudo-Goldstone (axion) direction.}
        \label{fig:Higgs2}
\end{figure}

More specifically, one may consider the following potential on the unit disk as a toy model for natural inflation
\be
V(Z, \bar Z) =V_0\left[ \left(1-  c^{-2}Z\bar Z\right)^{2} + A\, Z\bar Z\, \Bigl(Z^n+\bar{Z}^n + (Z\bar Z)^{n/2}\Bigr)\right]~,
\ee
with $c < 1$ and $0<A\ll 1$. In terms of $\vp$ and $\theta$, the potential is given by
\be
V(\vp, \theta)=V_0 \left[\left(1-c^{-2}\tanh^{2} \frac{\vp}{\sqrt{6\alpha}} \right)^{2}+ 4A\cos^{2}  {n\theta\over 2} \tanh ^{n+2} \frac{\vp}{\sqrt{6\alpha}}  \right]~.
 \label{toymodelpotential2}\ee
The term breaking the rotational symmetry was chosen so that for $A>0$ it is positive and monotonically grows with $\vp$ everywhere except $\theta =  {(2k+1)\pi\over n}$, where $k, n$ are integers, and $n > 1$. The shape of this potential is shown in Fig. \ref{fig:Higgs2} for $A \ll1$ and $n = 1, 2$ and  $8$.

\vskip 5pt

If $A$ is sufficiently small, and $1-c\ll 1$, the potential has  a flat direction at $\tanh  \frac{\vp}{\sqrt{6\alpha}} = c$, with a periodic potential
\be\label{natpot}
V(\theta) = 4\,V_{0}A\, c^{{n+2}}\cos^{2}  {n\theta\over 2} \ .
\ee

Naively,  one could expect that the circumference of the circle with any particular value of the field $\vp$  is given by $L=2\pi \vp$. However, because of the hyperbolic geometry reflected in the kinetic term \rf{kin}, the  circumference of the red circle shown in Fig. \ref{fig:Higgs} in terms of the canonically normalized angular field is greater by the  factor  
$ \sqrt{\frac{3\alpha}{2}} \sinh \left(\sqrt{\frac{2}{3\alpha}}\vp\right)$. Using equation \rf{minimum}, one finds that the circumference of the red circle is given by
\begin{equation}\label{kin2}
L =2\pi { \sqrt {6\alpha}\,  c \over 1-c^{2}} \ .
\end{equation}
Thus for $1-c\ll 1$ one finds the potential where the length of the Goldstone direction is extremely large.
This is the key feature of similar potentials to be considered in this paper, which allows to implement natural inflation scenario in this context.

Usually one represents the length of the circle as $L = 2\pi f_{a}$, where $f_{a}$ is the axion decay constant. However, the possibility to implement the natural inflation scenario depends not on the full length of the circle, but on the distance between the two nearby minima,
\begin{equation}\label{kin2n}
L_{n} = {2\pi\over n} { \sqrt {6\alpha}\,  c \over 1-c^{2}} \ .
\end{equation}
Therefore in this paper we will use an alternative definition of the axion decay constant, such that $L_{n} = 2\pi f_{a}$, where
\begin{equation}
 f_{a}  = \frac{\sqrt{6\alpha}\, c}{n(1-c^2)} \ .
 \label{beta}
\end{equation}
Natural inflation requires that the absolute value of the mass squared of the field $\theta$ at its maximum at $\theta = 0$ should be smaller than its height. This implies that natural inflation may occur only for $f_{a} > 1/2$.  To give a particular example, this condition can be met for $\alpha = 1/3$, $0.5 \lesssim c < 1$.

\section{Natural inflation in half-plane variables: E-models}
\label{sec:E}
Following \cite{Yamada:2018nsk}, we will  use the  K\"ahler potential and superpotential\footnote{Although these are different fields and parameters, in this section we abuse notation and use the variables $S,\ \varphi, \ \theta, \ A$ and $c$ again. This turns out to be convenient for describing the phenomenology for both models in Section \ref{sec:pheno}.}
\be
K=-3\alpha\log(T+\bar T)+S + \bar S +
G_{S\bar S}S\bar S \ , \qquad  W=W_0 \ .  \label{KE}
\ee
Here
\be
G_{S\bar S}=\frac{|W_0|^2}{(T+\bar T)^{3\alpha}V(T,\bar T)+3|W_0|^2(1-\alpha)} \ ,\label{GE}
\ee
and $T$ is the half-plane variable, $T = e^{-\sqrt{2\over 3 \alpha}\vp} + i\theta$, where $\vp$ is a canonically normalized inflaton field.
The kinetic term is given by
\begin{equation}\label{kinT2}
3\alpha\frac{\partial T \partial \bar{T}}{(T+\bar{T})^2} = \frac{1}{2}\left(\partial \vp\right)^2 + \frac{3\alpha}{4} e^{2\sqrt{\frac{2}{3\alpha}}\vp} \left(\partial \theta\right)^2 \ .
\end{equation}
The kinetic coefficient of $\theta$ plays an important role as in the T-model case.
\begin{figure}[h!]
\centering
		 \includegraphics[width=0.54\textwidth]{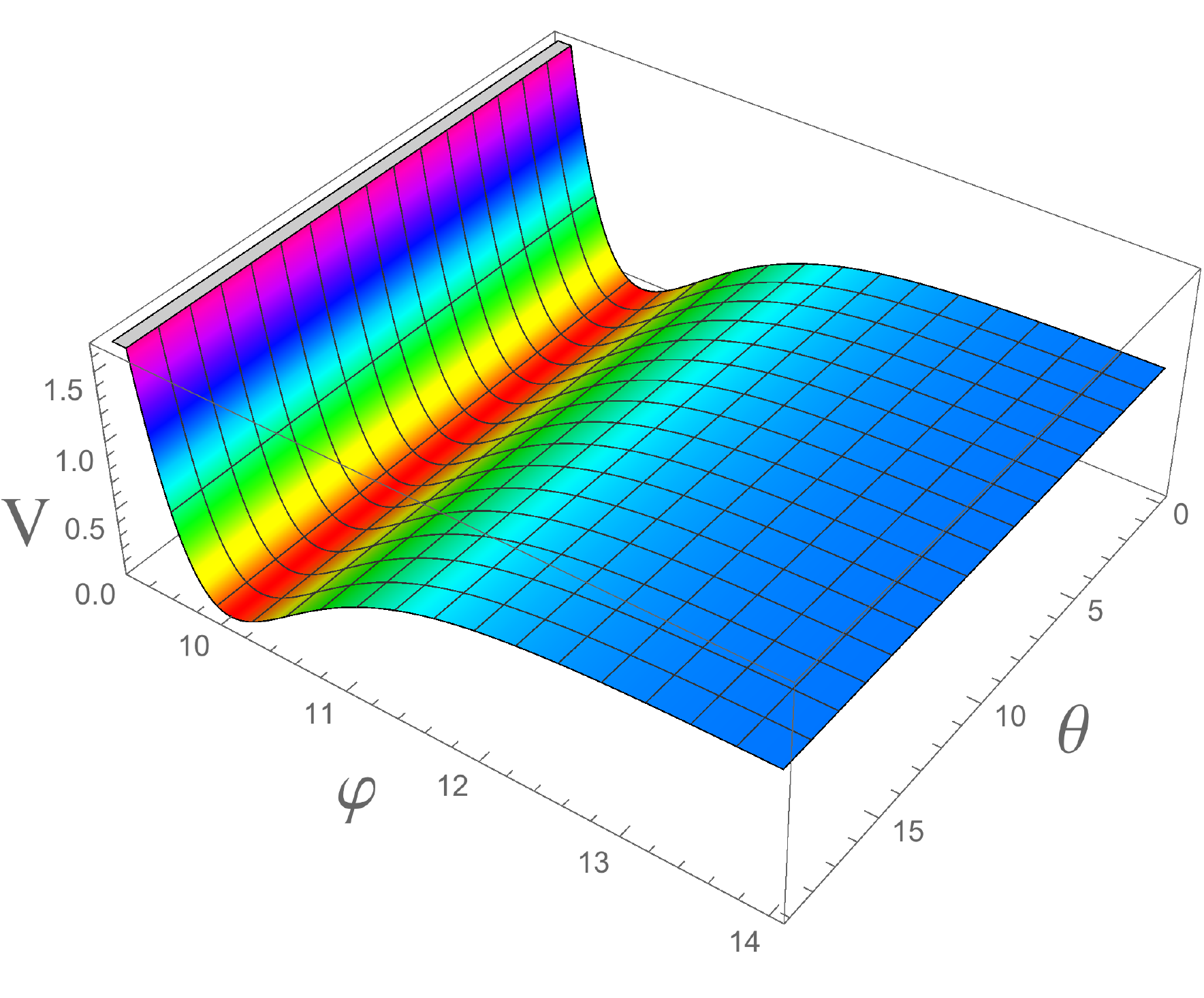}
        \caption{The shape of the E-model $\alpha$-attractor with a flat axion valley  at $\vp = c$, for $c=10$ and $\alpha = 1/3$.}
        \label{fig:E1}
\end{figure}

As a starting point, we will consider a potential
\be\label{Eflat}
V=V_{0}\Big(1 - {T + \bar T\over 2}\, e^{\sqrt{2\over 3 \alpha}c}\Big)^{2} = V_{0}\Big(1- e^{-\sqrt{2\over 3 \alpha}(\vp-c)}\Big)^{2} \ .
\ee
This is an E-model $\alpha$-attractor potential with respect to the field $\vp$, which has a flat axion direction at $\vp = c$,  see Fig. \ref{fig:E1}.

\begin{figure}[h!]
\centering
		 \includegraphics[width=0.55\textwidth]{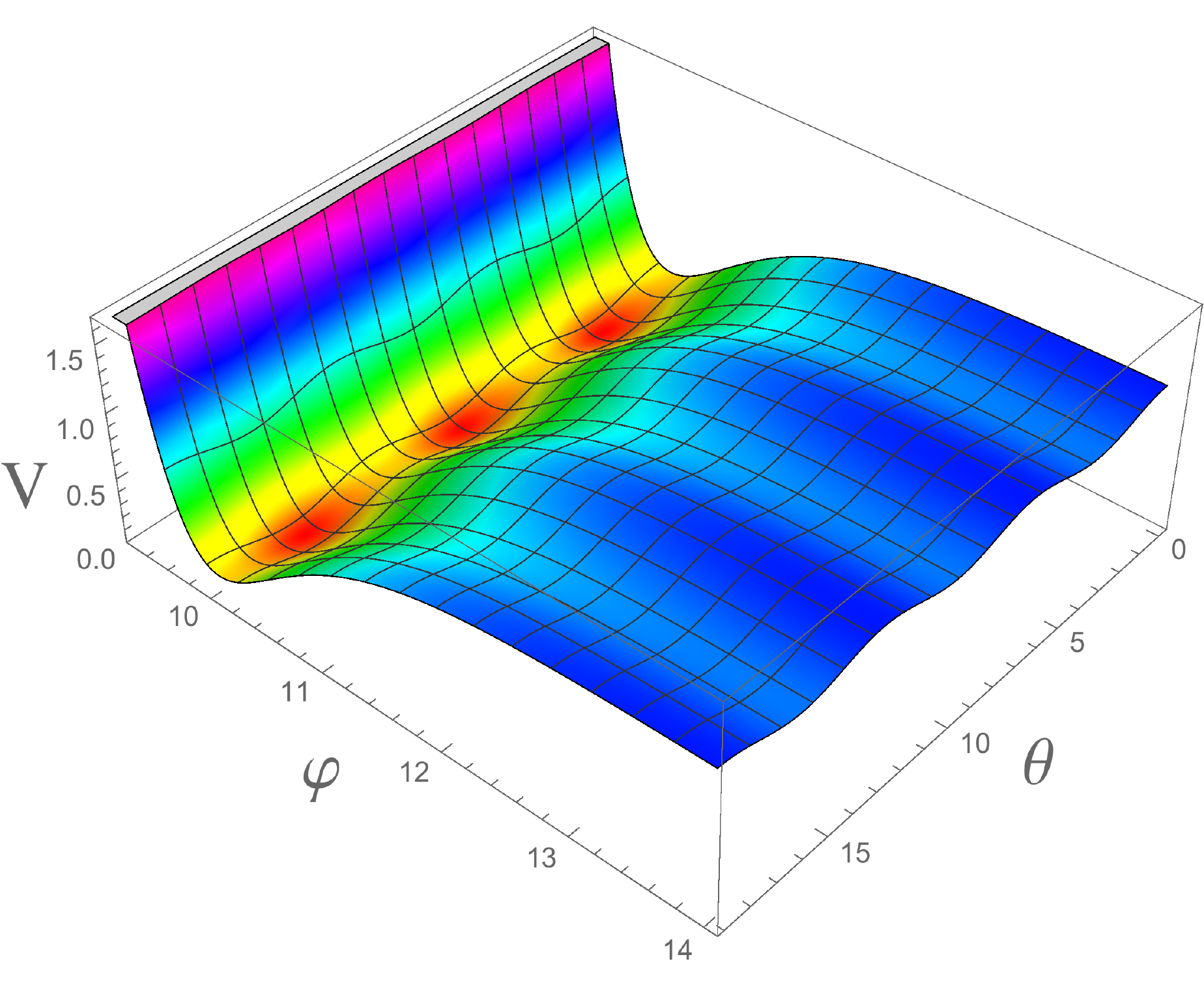}
        \caption{The shape of the E-model $\alpha$-attractor with a modulated axion valley  at $\vp =c$ and a natural inflation potential for the field $\theta$,  for $c=10$ and $\alpha = 1/3$.  }
        \label{fig:E2}
\end{figure}

To describe natural inflation, one may add to \rf{Eflat} a term   $4V_{0}A \cosh^{2} {n(T- \bar T)\over 4} $.
The resulting potential shown in see Fig. \ref{fig:E2} is given by
\be\label{Eax}
V=V_{0}\Big(1- e^{-\sqrt{2\over 3 \alpha}(\vp-c)}\Big)^{2} +4 V_{0 }A\,  \cos^{2}{n\,\theta\over 2} \ .
\ee
At large $\vp$, the last term in this expression coincides with the last term in \eqref{toymodelpotential2}.

This potential is periodic in $\theta$, with the period ${2\pi\over n}$, as in the  T-models considered earlier. The physical distance between the two nearby minima of the periodic potential is given by
\be
L_{n} = {2\pi\over n} \sqrt{3\alpha\over 2}\, e^{\sqrt{\frac{2}{3\alpha}}c} \ .
\ee
Thus for $c \gg \sqrt{3\alpha \over 2}$, the physical distance between the minima becomes exponentially large, and  one can have natural inflation driven by the field $\theta$.

We define the decay constant as
\begin{equation}
f_a={1\over n}\sqrt{\frac{3\alpha}{2}}e^{\sqrt{\frac{2}{3\alpha}}c},
\end{equation}
and at $\vp=c$, the canonically normalized axion $\hat\theta$ has the potential
\begin{equation}\label{natpot2}
V= 4  A\,\cos^{2}\left(\frac{\hat\theta}{2f_a}\right).
\end{equation}

 \section{Equations of motions and the slow-roll regime
 }\label{EOM}
Inflation in this model consists of two stages. First, the field $\vp$ rolls down from its large values, whereas the field $\theta$ remains nearly constant, in accordance with \cite{Achucarro:2017ing}. Then when the field comes close to the minimum, the field $\theta$ begins to evolve, and a stage of natural inflation may occur. The details of the process depend on initial conditions and on the parameters of the model. We performed a   full analytical and numerical analysis for the models discussed in the present paper. Fortunately, the main conclusions can be understood using the simple arguments following  \cite{Achucarro:2017ing}, which we present in this section.

Here we begin with the action of the scalar field $Z$ in T-models
\begin{equation}
\int d^4x\sqrt{-g}\left(-\frac12\partial^\mu\vp\partial_\mu\vp-\frac{3\alpha}4\sinh^2 \left(\sqrt{\frac{2}{3\alpha}}\vp\right)\partial_\mu\theta\partial^\mu\theta-V(\vp,\theta)\right)~,
\end{equation}
where we have used the parametrization  $Z=e^{i\theta}\tanh\frac{\vp}{\sqrt{6\alpha}}$.
The equations of motion for the homogeneous fields $\vp$ and $\theta$ in this theory look as follows:
\be\label{s2a}
\ddot\vp + 3H\dot \vp - \frac{1}{2}\sqrt{3\alpha\over 2}\, \sinh   \Bigl({2\sqrt{\frac{2}{3\alpha}}\vp}\Bigr) \, \dot\theta^{2} =   - V_{\vp}\ ,
\ee
\be\label{s1}
\ddot\theta+3H\dot\theta + 2\sqrt{\frac{2}{3\alpha}}\, {\dot\vp\, \dot\theta\over \tanh\sqrt{\frac{2}{3\alpha}}\vp}   =  -  {\frac{2}{3\alpha}}{V_{\theta}\over \sinh^{2} \sqrt{\frac{2}{3\alpha}}\vp} \ ,
\ee
where we have used the flat FLRW metric $g_{\mu\nu}=\text{diag}(-1,a^2(t),a^2(t),a^2(t))$, $H\equiv \dot a/a$ and the dot denotes a time derivative.

We will study the evolution of the fields during inflation at asymptotically large $\vp$, where the terms $ \frac{2V_\theta}{3\alpha\sinh^2 \left(\sqrt{\frac{2}{3\alpha}}\vp\right)}$ and $V_{\vp}$ in the right hand sides of these equations are exponentially small. An investigation of the combined evolution of the fields $\vp$ and $\theta$ in this case shows that if the evolution begins at sufficiently large values of the field $\vp$ in the region with $V >0$, then after a short period of relaxation, the fields approach a slow roll regime where the kinetic energy of both fields is much smaller than $V$. Under this assumption, one can neglect the first and the third terms in equation (\ref{s1}) and the first term in equation (\ref{s2a}). Taking into account that  $\tanh  \sqrt{\frac{2}{3\alpha}}\vp  \approx 1$ and $\sinh \left(2\sqrt{\frac{2}{3\alpha}}\vp\right) \sim 2 \sinh^{2} \left(\sqrt{\frac{2}{3\alpha}}\vp\right)\sim  {1\over 2}\, e^{2\sqrt {2\over {3\alpha}}\varphi}$ at $\vp \gg \sqrt{\alpha}$, one can represent equations for the fields $\theta$ and $\vp$ in the slow roll approximation as follows:
\be\label{ts1}
3H\dot \vp =  \frac{1}{4} {\sqrt {3\alpha\over 2}}\,e^{2\sqrt {2\over {3\alpha}}\vp} \, \dot\theta^{2} - V_{\vp}\ ,
\ee
\be\label{o1}
3H\dot\theta = -{8\over 3\alpha}\, e^{-2\sqrt {2\over 3\alpha}\vp} \ V_{\theta} \ ,
\ee
where $H^{2} = V(\vp,\theta)/3$. Using the last of these two equations, the first one can be simplified:
\be\label{simp}
3H\dot \vp = {8\sqrt 2\over 9\alpha\sqrt{3\alpha}}{ {V^{2}_{\theta}} \over  V}\, e^{-2\sqrt {2\over 3\alpha}\vp} - V_{\vp}\ .
\ee

Note that in the $\alpha$-attractor scenario the potential $V$ is a function of $\tanh  {\frac{\vp}{\sqrt{6\alpha}}}$, which behaves as $1-2 e^{-\sqrt {2\over 3\alpha}\vp}$ at large $\vp$. As a result, its derivative $V_{\vp}$ is  suppressed by the factor $e^{-\sqrt {2\over 3\alpha}\vp}$.
Consider for example the theory  \rf{toymodelpotential2}. In this model, at $\vp \gg \sqrt { \alpha}$, one has $V_{\theta} \sim V_{0}$, and $V_{\vp} \sim  V_{0}e^{-\sqrt {2\over 3\alpha}\vp}$, up to some factors depending on $A$, $c$, and $\theta$. Therefore the second term in \rf{simp} is suppressed by an extra factor $e^{-\sqrt {2\over 3\alpha}\vp}$ as compared to $V_{\vp}$, so it can be neglected,
and one can write the last equation at large $\vp$ as follows:
\be\label{o2}
3H\dot \vp =  - V_{\vp}\ .
\ee
 
  Comparing \rf{o1} and \rf{o2} one finds that during the early stage of inflation the value of  $\dot\theta$ is suppressed by the exponentially small  factor $e^{-\sqrt {2\over 3\alpha}\vp}$ with respect to $\dot\vp$.
In other words, in the regime  $\vp \gg \sqrt {3\alpha\over 2}$,
the field evolution is almost straight in the radial direction, {\it i.e.} it can be ``rolling on the ridge", despite of the angular dependence of the potential \cite{Achucarro:2017ing}.

An analogous result is valid for inflation in multi-field E-models.
The action of the scalar fields is
\begin{equation}
\int d^4x\sqrt{-g}\left(-\frac12\partial^\mu\vp\partial_\mu\vp-\frac{3\alpha}{4}e^{2\sqrt{\frac{2}{3\alpha}}\vp}\partial_\mu\theta\partial^\mu\theta-V(\vp,\theta)\right) \ .
\end{equation}
The equations of motion for the homogeneous fields $\vp$ and $\theta$  are
\begin{equation}
\ddot\vp+3H\dot\vp-\sqrt{\frac{3\alpha}{2}}e^{2\sqrt{\frac{2}{3\alpha}}\vp}\, \dot\theta^{2}+V_\vp=0 \ ,
\end{equation}
\begin{equation}
\ddot\theta+3H\dot\theta+2\sqrt{\frac2{3\alpha}}\dot\vp\dot\theta+\frac{2}{3\alpha}e^{-2\sqrt{\frac{2}{3\alpha}}\vp}V_\theta=0 \ .
\end{equation}

At large $\vp \gg \sqrt{\alpha}$ in the slow-roll approximation one has
\begin{equation}\label{Es1}
3H\dot\vp-\sqrt{\frac{3\alpha}{2}}e^{2\sqrt{\frac{2}{3\alpha}}\vp} \dot\theta^{2}+V_\vp=0 \ ,
\end{equation}
\begin{equation}\label{Es2}
3H\dot\theta+\frac{2}{3\alpha}e^{-2\sqrt{\frac{2}{3\alpha}}\vp}V_\theta=0 \ .
\end{equation}

These two equations coincide with equations \rf{o1}, \rf{simp} up to factors $\mathcal{O}(1)$.  Just as in the T-models, equation \rf{Es1} can be approximated by \rf{o2}, and the value of $\dot \theta$ typically is suppressed by the factor of $e^{-\sqrt {2\over 3\alpha}\vp}$ with respect to $\dot\vp$.
Therefore the field evolution in E-models with $\theta$-dependence  also shows the "rolling on the ridge" behavior at $\vp \gg \sqrt{\alpha}$.
This fact is responsible for the main result derived in  \cite{Achucarro:2017ing} for  T-models:  general inflationary predictions of the new class of models coincide with general predictions of single field $\alpha$-attractors for large number of e-foldings $N$:
\be
 n_s= 1-{2\over N}\ , \qquad r= {12\alpha\over N^2} \ .
\ee
The analysis performed above shows that the same conclusion should be valid for E-models as well.

For a better understanding of this result, one should take into account that in the simplest $\alpha$-attractor models the exponential suppression factor described above is given by
\be
e^{-\sqrt {2\over 3\alpha}\vp} \sim {3\alpha \over 8 N} \ ,
\ee
where $N$ is the number of remaining e-foldings of inflation at the time when the inflaton field is given by $\vp$. For example, for $\alpha = 1/3$ and $N \sim 55$, this suppression factor is about $2\times 10^{{-3}}$, which explains why the field $\theta$ practically does not move at the early stages of inflation.

However, one can always construct  models where some of the assumptions of our analysis are not satisfied.  This may happen, in particular, in the models where inflation in the $\alpha$-attractor regime driven by the field $\vp$ is followed by another stage of inflation, such as  the natural inflation scenario driven by the angular component of the inflaton field. We will discuss this possibility now.

\section{Phenomenology}\label{sec:pheno}

In this section we describe the phenomenology of our toy models for natural inflation. The phenomenology of natural inflation in the E-model is very similar to that of the $U(1)$ symmetric T-model, and our results are phrased in such a way that they apply to both. 

Let us first study the simplest $U(1)$ symmetric models with $A= 0$. In that  case, after a long stage of inflation, the value of $\dot \theta$ vanishes, and inflation is driven by the field $\vp$ \footnote{It is straightforward to verify that this regime does not suffer from geometric destabilization, see eq. (B.8) in \cite{Achucarro:2017ing}.}. After the end of inflation, the  field $\vp$ falls to the minimum of its potential (the Goldstone valley with respect to the field $\theta$), and oscillates there, which leads to reheating.  In that case, the observational predictions of inflation coincide with the prediction of the single field $\alpha$-attractors, despite the existence of the massless particles $\theta$ produced by reheating.

Clearly, for sufficiently small values of $A$ these conclusions will remain valid, unless the stage of $\alpha$-attractor inflation driven by the field $\vp$ is followed by another stage of inflation involving the field $\theta$. For example, after the stage of oscillations, the field $\vp$ may settle near the minimum of its potential, but the field $\theta$ may be away from its minimum. Then the field $\theta$ will start moving, driving a stage of natural inflation in the potential \rf{natpot}, \rf{natpot2}.

As we already mentioned, natural inflation is possible only for $f_{a} > 1/2$. Thus for sufficiently small $A$ and $f_{a} < 1/2$ we have only one stage of inflation. It is driven by the field $\phi$ and leads to the universal predictions
 \begin{equation}
  n_s = 1-\frac{2}{N}, \quad r = \frac{12\alpha}{N^2} .
  \label{universalpredictions}
 \end{equation}

Meanwhile for $f_{a} \gg 1$, the $\alpha$-attractor regime can be followed by the stage of natural inflation.   In that case, the results depend on the duration of this stage. If this stage is longer than $N \sim 55$ e-foldings, all observational predictions will be determined by the stage of natural inflation, and the first stage driven by the field $\vp$ will be forgotten.  As we will show in the Appendix, this regime occurs for $f_{a} \gg \sqrt{N}/\pi \sim 3$.  In that case, if  we start with a typical initial value of $\theta$ in the range $|\theta| \sim \pi$, the stage of natural inflation will last more than $N \sim 55$ e-foldings.  Thus we have found a rather non-trivial realization of natural inflation in the new context provided by the theory of $\alpha$-attractors. The latest observational data do not offer much support to natural inflation as compared to the simplest $\alpha$-attractor models, but more data are needed to reach a definite conclusion  \cite{Ade:2015lrj, Ade:2015ava}.

Finally, there is an intermediate regime $1/2\lesssim f_{a} \lesssim \sqrt{N}/\pi \sim 3$, where the duration of natural inflation typically is smaller than $N \sim 55$ e-foldings. In this regime, the large scale CMB anisotropy will be described by the universal $\alpha$-attractor results \ref{universalpredictions}. However, in these results instead of the full number $N \sim 55$ one should use $N_{\alpha}=N_{total}-N_{{natural}}$, where $N_{total} \sim 55$, and $N_{{natural}}$ is the total number of e-foldings during the stage of natural inflation. Unless $N_{\alpha} \gg  N_{{natural}}$, the corresponding predictions will lead to unacceptably small values of $n_{s}$.

From this perspective, more complicated models  studied in this paper do not offer any advantages as compared to the simplest $\alpha$-attractor models. Nevertheless, we believe that it is important that natural inflation is indeed possible in a broad class of models based on supergravity, including the T-models with the Mexican hat potentials. Moreover, as we will explain in the next section, the new methods of inflationary model construction described above may be helpful for development  of a more general class of inflationary models.

\section{Multifield evolution for large $A$}

\begin{figure}[h!]
\centering
		\includegraphics[width=0.55\textwidth]{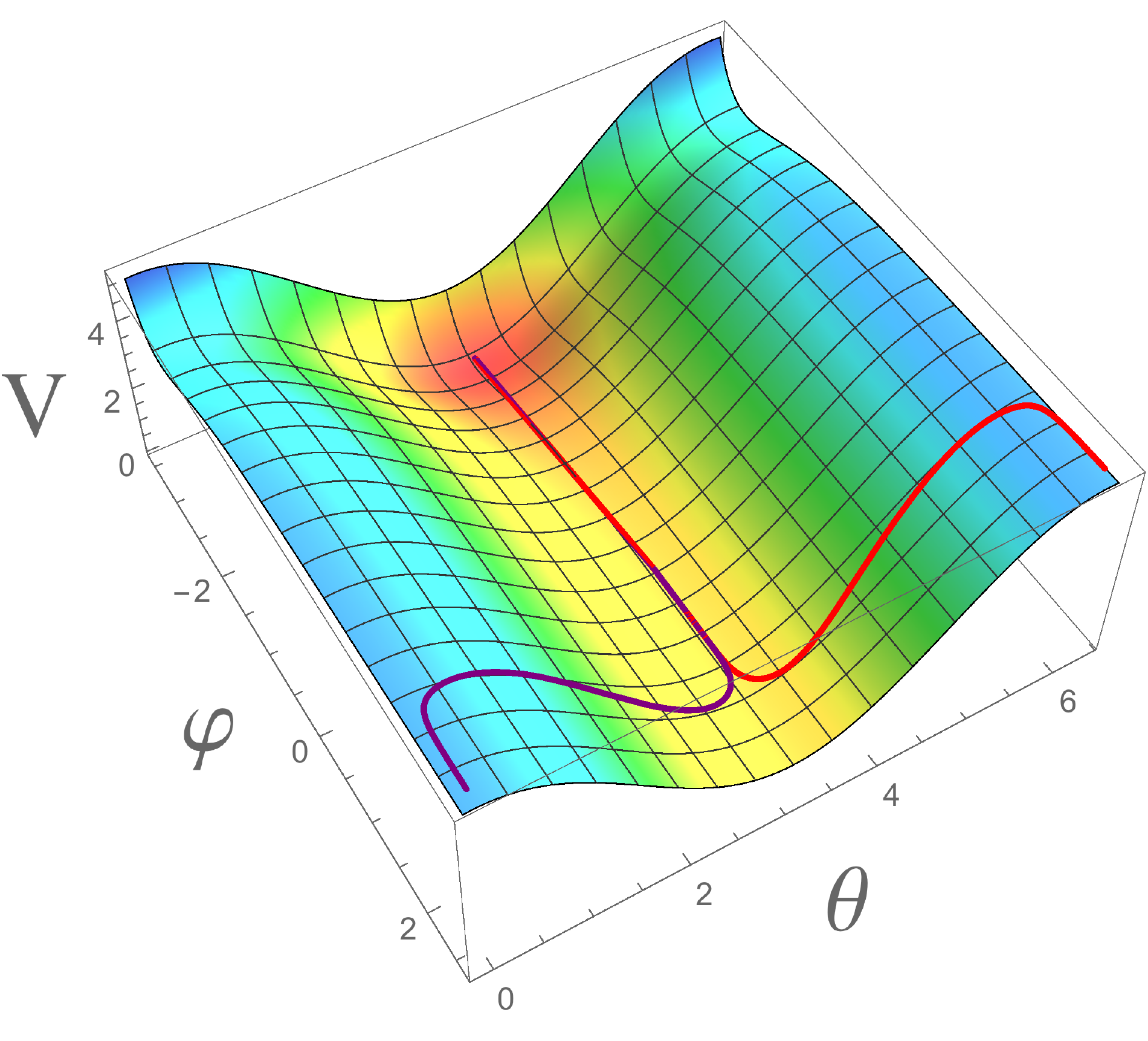}
        \caption{A model with two different $\alpha$-attractor stages  separated by an intermediate inflationary regime driven by a combined motion of the fields $\vp$ and $\theta$.}
        \label{fig:E2u}
\end{figure}

In the analysis of phenomenology, we studies only the small $A$ case. In the situations where the amplitude $A$ of the modulations of the potential is large, the inflationary evolution can be quite complicated. One of the examples is shown in Fig. \ref{fig:E2u}. We considered the theory \rf{Eax} with $\alpha = 1/3$,  $A = 1$ and  $c  = -3$, and plotted the height of the potential in units of $V_{0}$.  Note that the minimum of the potential is at $\vp = -3$.

Initially the field $\vp$ was rolling along the ridge of the potential, as shown either by a red line, or a purple line. The field $\theta$ was not evolving because of the  factor  $e^{-\sqrt {2\over 3\alpha}\vp}$ in its equations of motion suppressing its evolution. However, when the field $\vp$ decreases, the factor  $e^{-\sqrt {2\over 3\alpha}\vp}$ decreases, and the field $\theta$ rapidly falls down. Because of the interaction between the derivatives of $\vp$ and $\theta$ in the equations of motion of these fields, the field $\vp$
moves back a bit. This unusual effect is similar to the behavior found in a closely related context in  \cite{Kallosh:2014qta,Christodoulidis}. When the field $\theta$ reaches the minimum of its potential at $\theta = \pi$, a new stage of inflation driven by the field $\vp$ begins.  If this stage is long enough (as is the case for the model described here), the observational data will be determined by this last stage of $\alpha$-attractor inflation.

\begin{figure}[h!]
\centering
		 \includegraphics[width=0.55\textwidth]{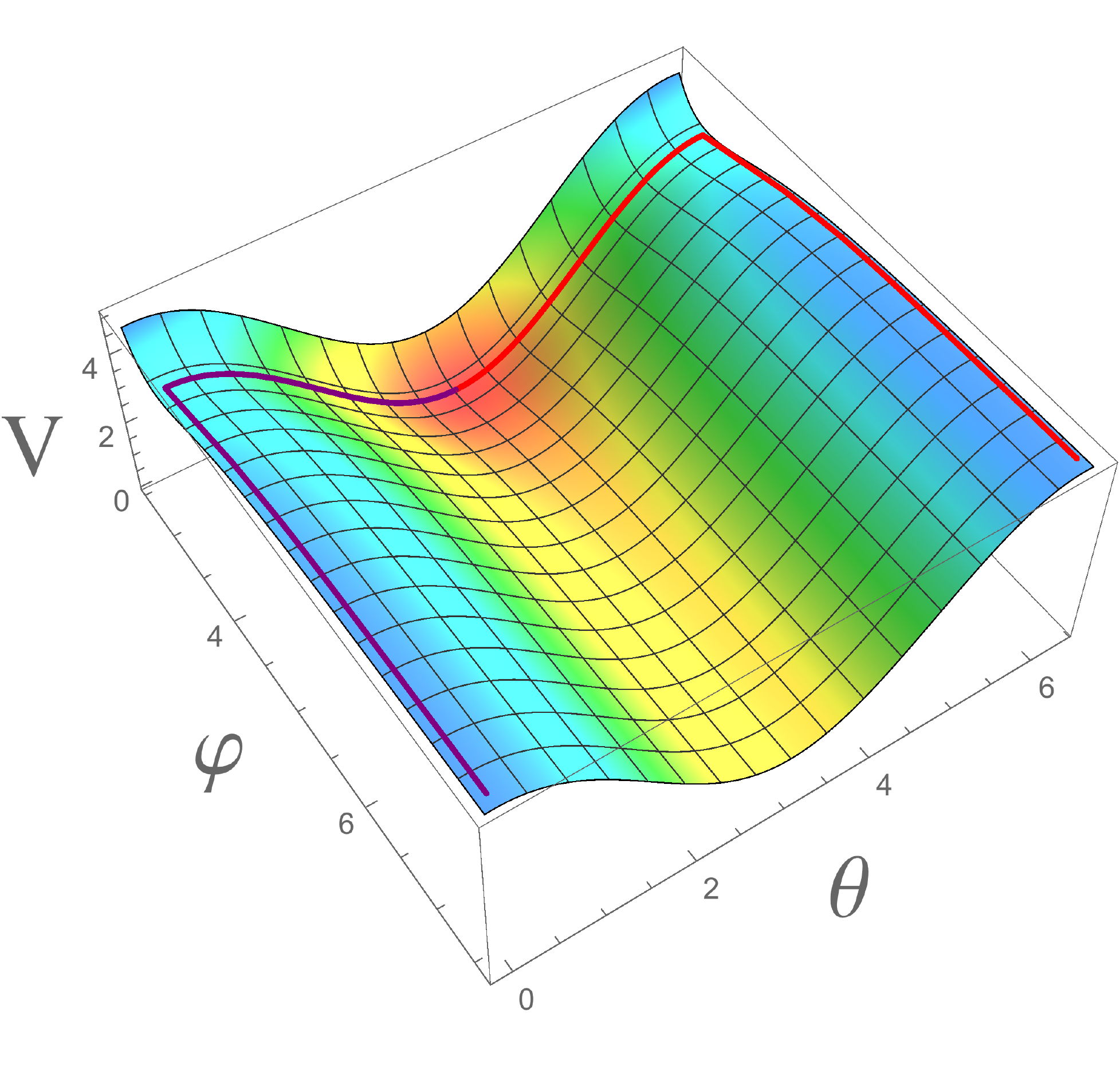}
        \caption{$\alpha$-attractor inflation along each of the ridges of the potential, shown either by a red line, or a purple line,  follows by natural inflation driven by the field $\theta$.}
        \label{fig:E2v}
\end{figure}

On the other hand, if we take the same potential with $c = +3$, the first stage of $\alpha$-attractor inflation will continue for a long time, until the field $\vp$ falls down to its minimum at $\vp = c$, and then the new stage of inflation begins, which  will be driven by the field $\theta$ with the natural inflation potential, see Fig. \ref{fig:E2v}.

\section{Models with a finite plateau}\label{singul}

Until now we studies the standard $\alpha$-attractor models with nonsingular potentials in terms of the original geometric variables. However, one of the way to describe the E-models such as the Starobinsky model is to start with a T-model and assume that the original potential is singular in one of the two directions  \cite{Kallosh:2015zsa}. Let us see what may happen in the context of the $U(1)$-symmetric models discussed in this paper.

\begin{figure}[h!]
\centering
		 \includegraphics[width=0.55\textwidth]{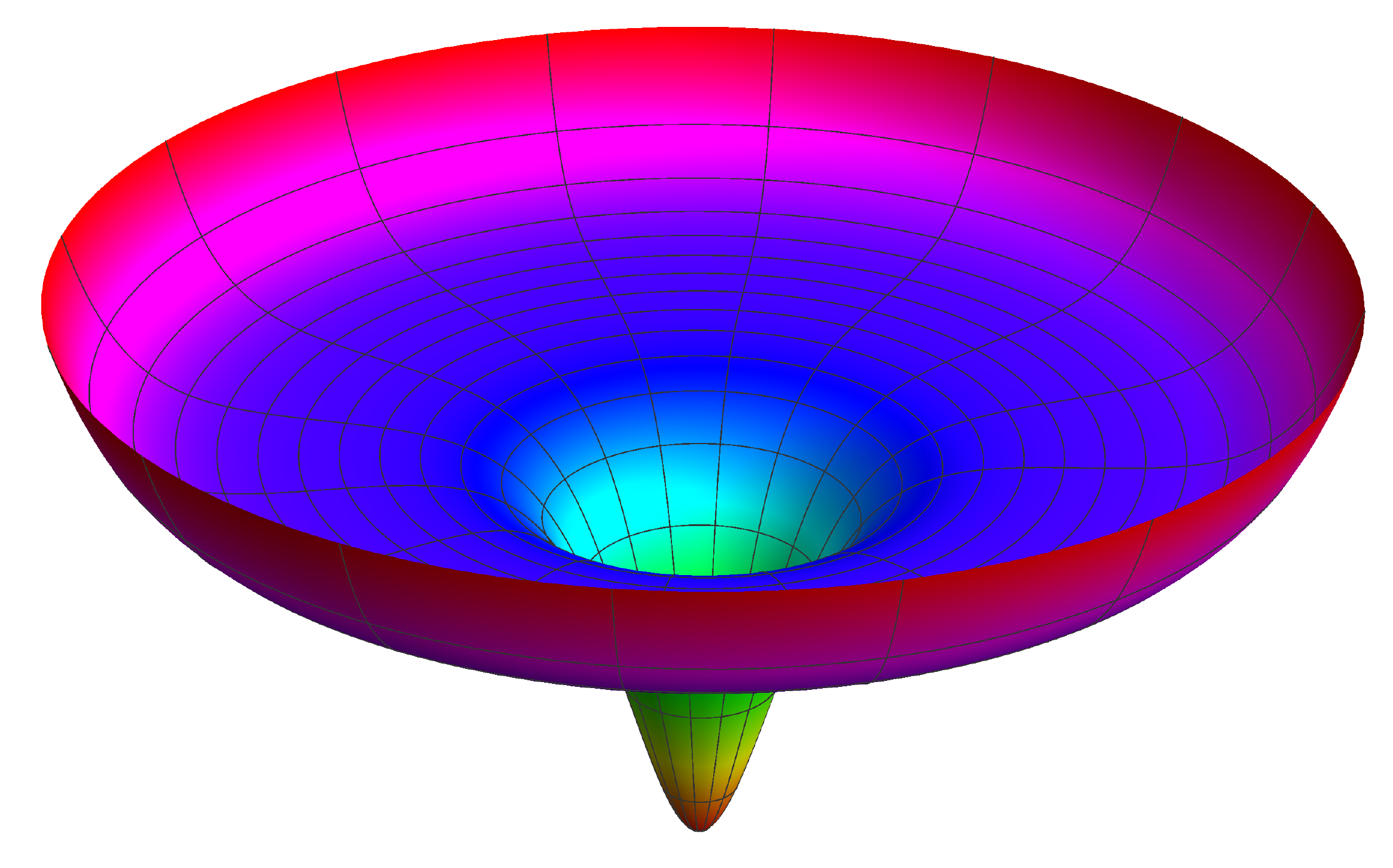}
        \caption{A model with an $\alpha$-attractor  plateau potential bounded by an exponentially steep wall, which emerges because of the singularity of the potential \rf{sing} at $|Z| = 1$. }
        \label{fig:E2w}
\end{figure}

As an example, consider the T-model scenario discussed in Section \ref{sT}, with
\be\label{sing}
V(Z, \bar Z) =V_0 \ Z\bar Z \, \left(1+ {A\over 1- Z\bar Z}\right).
\ee
If the parameter $A$ is very small, the last term is important only in the vicinity of the singularity. The potential in terms of the inflaton field $\vp$ looks as follows:
\be
V(\vp, \theta)=V_0 \left( \tanh^{2} \frac{\vp}{\sqrt{6\alpha}}  +  A\sinh^{2}   \frac{\vp}{\sqrt{6\alpha}} \right)~.
 \label{toymodelpotentialhyper}
 \ee
This function looks strikingly similar to some of the potentials studied in the previous sections, and yet it is very much different, see Fig. \ref{fig:E2w}.

At the first glance, this potential looks like a good candidate for the role of a potential supporting hyperinflation \cite{Brown:2017osf}. A preliminary investigation of this issue indicates that it does not satisfy the required conditions formulated in \cite{Mizuno:2017idt}, though one might achieve this goal by modifying the nature of the singularity of the potential.

However, the structure of the potential \rf{toymodelpotentialhyper} suggests another interesting possibility.  At large $\vp$ this potential has asymptotic behavior
\be
V(\vp, \theta)=  {V_0\, A\over 4} \, e^{{\sqrt{2\over 3\alpha}}\vp}~.
 \label{toymodelpotentialhyper2}
 \ee
Cosmological evolution in such potentials is well known  \cite{Liddle:1988tb}: 
\be\label{power}
a(t) = a_{0}\ t^{3\,\alpha} \ .
\ee
 For $\alpha = 1/3$ we have a regime with $a \sim t$, exactly at the boundary between the accelerated and decelerated expansion of the universe. In that case, the energy of the homogeneous component of the scalar field decreases as $a^{{-2}}$, i.e. much more slowly than the energy of dust $\sim a^{{-3}}$ and of the relativistic gas $\sim a^{{-4}}$.
This makes the solution of the problem of initial conditions proposed in \cite{Carrasco:2015rva,East:2015ggf} much simpler: If initially the energy of the homogeneous field was comparable to other types of energy, then in an expanding universe it gradually starts to dominate.  Meanwhile for  $\alpha > 1/3$ the power-law solution \rf{power} describes  inflation. It may begin already at the Planck density, which solves the problem of initial conditions in this class of models along the lines of \cite{Linde:1985ub}.

For $\alpha = 1/3$, this  potential  provides the power-law expansion with $a(t) \sim t$, exactly at the boundary between the accelerated and decelerated expansion of the universe.  In this regime, the energy density of the universe decreases as $1/a^{2}$, similar to what happens in the open universe scenario. This regime is very helpful for solving the problem of initial conditions for the subsequent stage of inflation supported by the plateau potential \cite{Carrasco:2015rva,East:2015ggf,Linde:2017pwt}.
Meanwhile for any value $\alpha > 1/3$ one can have inflation starting  at the Planck density, which immediately solves the problem of initial conditions for inflation in this scenario  \cite{Linde:2017pwt}.

Similar conclusions are valid not only for T-models, but also for E-models with singular potentials, or for the single-field $\alpha$-attractors with singular potentials \cite{Linde:2017pwt}. In particular, if one adds  to the E-model potential \rf{Eflat} a term  $\sim(T+\bar T)^{{-1}}$, which is  singular at $T \to 0$, the potential will also acquire an exponentially growing  correction $\sim e^{{\sqrt{2\over 3\alpha}}\vp}$, just as in the T-model discussed above.

Note that the possibility to have inflation at asymptotically large values of the fields in such models
does depend on $\alpha$ and on the behavior of the potential near the singularity. For example for the T-models with the singularity $(1- Z\bar Z)^{n}$, the inflationary regime at the asymptotically large $\vp$ is possible for $\alpha > n/3$.

 The scenario outlined above may have  non-trivial observational implications. The amplitude of scalar perturbations $A_{s}$ in the slow-roll approximation is given by 
 \be
 A_{s}  =  {V^{3}\over 12\pi^{2}V'^{2}} \ .
 \ee
 In the theory without the singular term (i.e. with $A = 0$), the amplitude grows monotonically at large $\vp$ since $V$ approaches a plateau and $V'$ decreases. Meanwhile in the theory with the potential \rf{toymodelpotentialhyper}, the value of $V'$ exponentially decreases, and therefore $A_{s}(\vp)$ reaches a maximum and then falls down exponentially at large $\phi$. This effect suppress the large wavelength amplitude of inflationary perturbations.  If the transition between the two inflationary regimes is sufficiently sharp (which depends on $\alpha$ and the nature of the singularity of the potential) and corresponds to  $N \sim 55$, this effect may help to suppress the low-$\ell$ CMB anisotropy \cite{Linde:1998iw,Contaldi:2003zv}.

\section{Other two-field attractors} 

\begin{figure}[h!]
\centering
		 \includegraphics[width=0.55\textwidth]{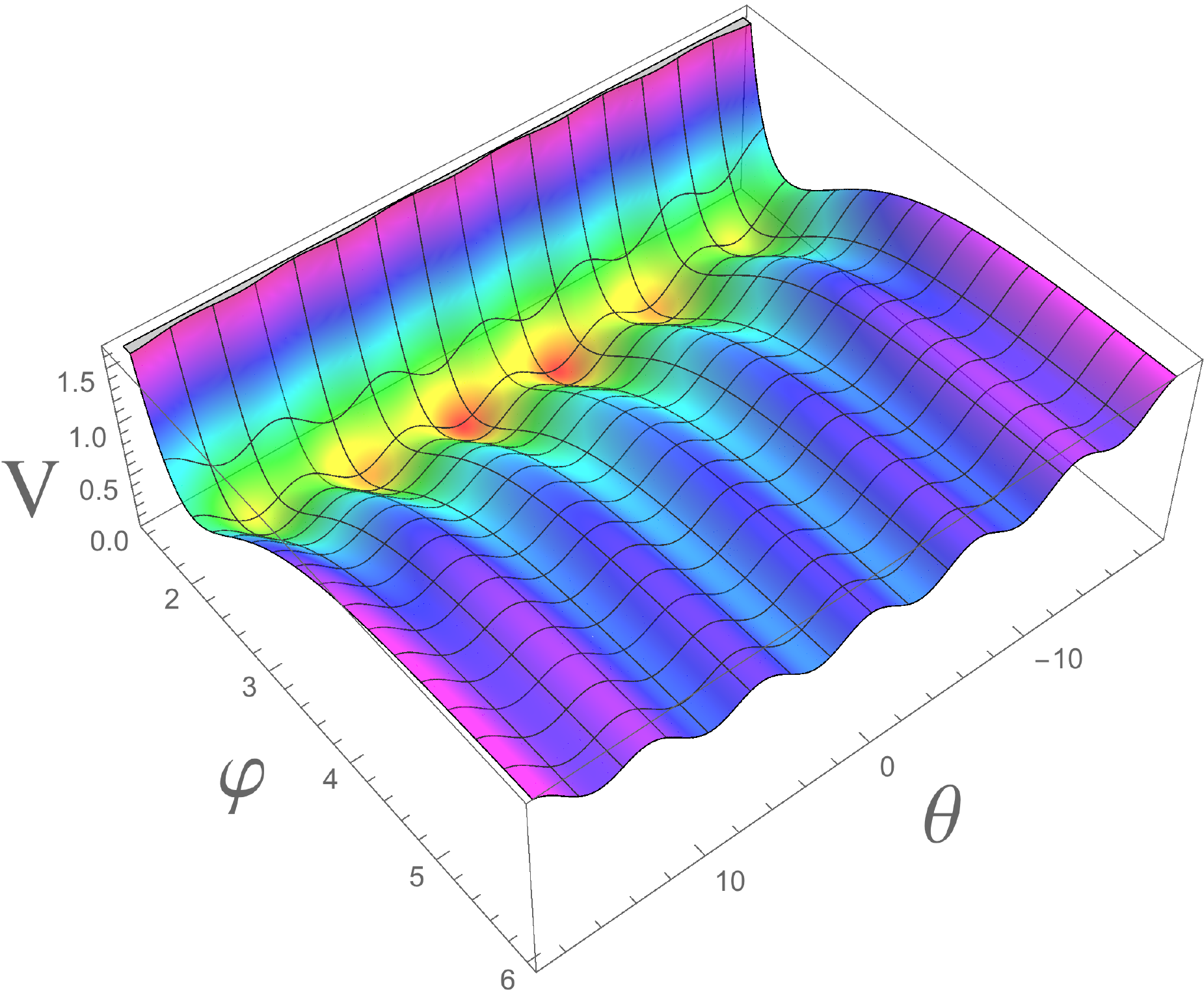}
        \caption{An example of the $\alpha$-attractor monodromy potential \rf{Eaxcc}.}
        \label{fig:E2pp}
\end{figure}

Finally, we should mention a more general class of models which can be constructed by our methods:  Instead of the term    $4A \cosh^{2} {n(T- \bar T)\over 4} $, one can add to \rf{Eflat} an arbitrary function $f\bigl[{1\over 2}((T+\bar T),-{i\over 2}(T-\bar T)\bigr]$, which leads to the potential
\be\label{Eaxa}
V=V_{0}\Big(1- e^{-\sqrt{2\over 3 \alpha}(\vp-c)}\Big)^{2} +f(\vp, \theta) \, .
\ee
In particular, for $f =- {m^{2}\over 8} (T-\bar T)^{2}$ one has a combination of the E-model $\alpha$-attractor potential and a quadratic potential,
\be\label{Eaxbb}
V=V_{0}\Big(1- e^{-\sqrt{2\over 3 \alpha}(\vp-c)}\Big)^{2} +{m^{2}\over 2}\theta^{2} \ ,
\ee
and for $f =- {m^{2}\over 8} (T-\bar T)^{2} + 4V_{0}A \cosh^{2} {n(T- \bar T)\over 4}$ one has a ``monodromy'' potential \cite{Silverstein:2008sg,McAllister:2008hb} shown in Fig. \ref{fig:E2pp}.
\be\label{Eaxcc}
V=V_{0}\Big(1- e^{-\sqrt{2\over 3 \alpha}(\vp-c)}\Big)^{2} +4 V_{0 }A\,  \cos^{2}{n\,\theta\over 2} +{m^{2}\over 2}\theta^{2} \ .
\ee

Thus  one can find a large class of consistent inflationary models without making a requirement that the axion field $\theta$ must be strongly stabilized during inflation.


\

\section{Conclusions}

In this paper we have shown how to realize $U(1)$ symmetry breaking and shift symmetric potentials in the framework of $\alpha$-attractors. This might lead to a novel realization of natural inflation in supergravity, in a similar way as originally proposed in \cite{Freese:1990rb} through non-perturbative instanton corrections.

To illustrate the phenomenology of these models of natural inflation, we consider a concrete example of a Higgs-type potential with a circular minimum at radius $|Z| = c$, together with sinusoidal corrections of amplitude $A$ and $n$ minima.

If we take a simple and natural value of $0<c<1$, not too close to 1, we recover the results we found in \cite{Achucarro:2017ing}. We find universal predictions fairly insensitive to modifications of the potential. This regime could be interesting for applications such as suppression of isocurvature perturbations in the axion dark mater \cite{Linde:1991km,Ema:2016ops,Yamada:2018nsk,Ema:2018abj,LindeYamada}.

If we fine-tune $1-c^2 \ll 1$, the circumference of the valley and the axion decay constant $f_{a}$ become very large because of the hyperbolic geometry.  This opens up the possibility of inflation that proceeds in two stages. For example, inflation may start in the  $\alpha$-attractors regime. Then, after the fields have reached the valley of the Higgs potential, the stage of natural inflation may begin.

For naturally small values of $A$ we find the following generic behavior.
\begin{itemize}
 \item If $f_a \gtrsim  \sqrt{N}$,  starting with a random initial angle we typically find the predictions of natural inflation with a large effective decay constant
\begin{equation}
\label{predictionsnaturalinflation}
  n_s = 1-\frac{2}{N}\ , \qquad r = \frac{8}{N} \ .
 \end{equation}
 Here $N$ is the number of efolds before the end of inflation where the observable scales cross the horizon.
 \item If $\frac{1}{2} \lesssim f_a \lesssim   \sqrt{N}$ or if we happen to start close to one of the  minima, we will find the $\alpha$-attractor predictions with a reduced number of efolds
\begin{equation}
   n_s \approx 1-\frac{2}{N-N_{II}}\ , \qquad r \approx \frac{12\alpha}{(N-N_{II})^2}\ ,
\end{equation}
where $N_{II}$ is determined by the initial angle. These predictions are shifted left upwards in the $(n_s, r)$ plane with respect to the predictions of $\alpha$-attractor.
\item In the simplest case  $f_a \lesssim \frac{1}{2}$ natural inflation does not occur, and the predictions coincide with the standard $\alpha$-attractor predictions
\begin{equation}\label{alp}
   n_s \approx 1-\frac{2}{N}\ , \qquad r \approx \frac{12\alpha}{N^2}\ .
\end{equation}

\end{itemize}

Our results extend to any potential with a flat minimum well in the $\alpha$-attractor regime, and any type of small angular correction breaking the shift symmetry. In particular, we find exactly the same predictions for natural inflation in the E-models.

The cosmological evolution in the models with very large $A$ sometimes involves a more complicated behavior.  However, in all cases where the last 50-60 e-foldings of inflation occur in the $\alpha$-attractor regime, the cosmological predictions are quite stable with respect to modifications of the inflaton potential and coincide with the universal single-field $\alpha$-attractor predictions \rf{alp},  in agreement with the general conclusions of \cite{Achucarro:2017ing}.

 The new class of models discussed in this paper allow various generalizations. In particular, in Section \ref{singul} we described a new class of models, where the potentials contain small terms which become singular at the boundary of the moduli space. Such models allow to describe $\alpha$-attractors with potentials having a plateau of a finite size. We found that in certain cases the exponentially rising potentials at asymptotically large values of the field $\vp$ can support inflation all the way up to the Planck density, which provides a simple solution to the problem of initial conditions for inflation in such models  \cite{Linde:2017pwt}. 

\

 \noindent{\bf {Acknowledgements:}}  We thank Renata Kallosh, Diederik Roest and Alexander Westphal for stimulating discussions and collaborations on related work.  The work  of AL and YY is supported by SITP and by the US National Science Foundation grant PHY-1720397.  DGW and YW are supported by a de Sitter Fellowship of the Netherlands Organization for Scientific Research (NWO). The work of AA is partially supported
by the Netherlands' Organization for Fundamental Research in Matter (FOM),  by the Basque Government (IT-979-16) and by the Spanish Ministry MINECO  (FPA2015-64041-C2-1P).


\section{\boldmath{Appendix A: Observational predictions as a function of $f_{a}$} }

\subsection{Parameter ranges}
\label{sec:parameterranges}
We shortly discuss the range of values the model parameters can take to which our generic results in Section \ref{subsec:singlefield} apply.

The parameter $A$ determines the size of the correction to the potential \rf{toymodelpotential2} and \rf{Eax}.
For our generic results to apply, we only need to ensure that we stay away from the multi-field regime in the valley of the potential. In case of the T-model, this leads to the constraint\footnote{\label{footnote}
In the slow roll regime we can roughly estimate the size of the effective mass $\mu$ of the radial perturbations as
$\frac{\mu^2}{H^2} \approx 3\frac{V_{\varphi\varphi}}{V}$. For simplicity we neglect the corrections from the curvature of the field space, which scale like $\epsilon/\alpha$, and the turn rate, which is smaller than $H$.
Using $\rho \approx c\sim 1$ this becomes
$\frac{\mu^2}{H^2} \approx \frac{6}{A f_a^2 \cos^2(n \theta/2)}$
for the T-model \rf{toymodelpotential2}.
As soon as this ratio becomes order one we have to be careful about the multi-field effects and perform a numerical analysis. To find the constraint on $A$ we first solve for the evolution of $\theta(N)$ in the gradient flow approximation to find
$\cos^2(n \theta/2) = 1 - e^{-n^2 N/f_a^2}$. 
In the regime $f_a^2 \gtrsim n^2 N$ the condition $\mu^2/H^2\gg 1$ translates to $A \ll \frac{6}{n^2 N}$. In the regime $f_a^2 \lesssim n^2  N$ we instead find $A \ll \frac{6}{f_a^2}$. Together they lead to the result quoted in \eqref{improvedconstraintA}. For the E-model \rf{Eax} we can estimate $\frac{\mu^2}{H^2} \approx \left|3\frac{V_{\vp\vp}}{V}\right|_{\vp=c} \approx \frac{1}{\alpha A\cos^2(n\theta/2)}.$
Hence, we need to have $\alpha A\cos^2(n\theta/2) \ll 1$. Therefore \eqref{EimprovedconstraintA} is a sufficient condition.}
\begin{equation}
 A \ll \frac{6}{n^2 N},
 \label{improvedconstraintA}
\end{equation}
with $N\sim 60$ the number of efolds between horizon crossing and the end of inflation. Similarly, for the E-model a sufficient condition is
\begin{equation}
\alpha A\ll 1.
\label{EimprovedconstraintA}
\end{equation}
In Section \ref{subsec:singlefield} 
we assume these conditions are obeyed and study the generic behavior of this system.

\subsection{Phenomenology in the effectively single field regime}\label{subsec:singlefield}
We assume that the condition \eqref{improvedconstraintA} is obeyed. We generically first have a stage of inflation in the two field $\alpha$-attractor regime until $\varphi$ reaches the valley of the symmetry breaking potential. Using the results of \cite{Achucarro:2017ing}, we know the system behaves effectively like the single field alpha attractor. We outlined the derivation of this result in Section \ref{EOM}. Consequently we enter a second stage of natural inflation in the bottom of the Higgs potential.

The value of the effective axion decay constant $f_{a}$ determines the phenomenology of these family of models. To see this, the phenomenologically distinct regimes we are interested in are:
\begin{enumerate}
 \item The natural inflation potential allows for slow roll inflation to happen: $\eta_\theta \ll 1$ at the top of the hill of the angular potential at the minimum of the Higgs potential.
 $$
 \eta_\theta \sim -\frac{V_{\theta\theta}/V}{\frac{3\alpha}{2}\sinh^2\sqrt{\frac{2}{3\alpha}}\varphi} \ll 1 \ .
 $$
 Natural inflation is possible for $ f_{a}\gg 1/2$.
 \item The stretching of the angular potential is large enough to allow for a quadratic expansion about the origin of the natural inflation potential in the observable range of scales:
 $$
 N/ f_\ast^2 \lesssim 1, \quad \text{or} \quad f_{a} \gtrsim \sqrt{N},
 $$
 with $N \sim 60$ the number of efolds before the end of inflation where the observable modes cross the horizon.
 \item The probability to have $N > 60$ e-folds of natural inflation starting with a random initial angle is close to 1 (assuming the quadratic expansion around the minima holds true):
 $$
 \theta_\text{can}(N) = \sqrt{4 N} \quad \longrightarrow \quad \theta(N) = \left.\frac{\sqrt{4 N}}{\sqrt{\frac{3\alpha}{2}}\sinh\sqrt{\frac{2}{3\alpha}}\varphi}\right|_{ f_{a}=f_\ast}.
 $$
 This leads to
 $$
 P \equiv 1 - \frac{\theta(N) }{2\pi c / n}= 1-\frac{\sqrt{N}}{ f_{a}\ \pi},
 $$
 with $2\pi c / n$ the distance between the minima. Therefore in order to have $P$ close to 1 we need the second contribution to be negligible
 $f_{a} \gg \frac{\sqrt{N}}{\pi}.$ This is automatically satisfied, since we already assumed the quadratic approximation.
\end{enumerate}
Therefore, for small enough $A$, the behavior of our system is very simple and splits in three regimes determined by the value of $ f_{a}$. It can be summarized as follows.
\begin{itemize}
 \item If there is no axion potential or if $ f_{a} \lesssim \frac{1}{2}$ we recover the universal predictions of $\alpha$ attractors \rf{universalpredictions}, as shown in  our previous paper
 \begin{equation}
  n_s = 1-\frac{2}{N}, \quad r = \frac{12\alpha}{N^2} .
  \label{universalpredictions2}
 \end{equation}
 \item  If $ f_{a} \gtrsim \sqrt{N}$, we would generically expect to have sufficient efolds in the valley of the potential where natural inflation takes place. We get the same predictions as natural inflation with an large effective axion decay constant, which coincide with those of chaotic inflation
 \begin{equation}
  n_s = 1-\frac{2}{N}\ ,  \qquad r = \frac{8}{N} \ .
  \label{naturalinflationpredictions}
 \end{equation}
 \item If $ f_{a}$ takes values in between, we most likely have not sufficient number of efolds in the regime of natural inflation. Therefore the predictions for the large-scale perturbations are the same as for the $\alpha$-attractor predictions, but with a reduced number of e-folds. If we denote $N_{II}$ as the number of e-folds in the natural inflation potential, we find
\begin{equation}
  n_s \approx 1-\frac{2}{N-N_{II}}, \quad r \approx \frac{12\alpha}{(N-N_{II})^2}.
  \label{alphaattractorpredictions}
 \end{equation}
This means that the predictions move left upwards in the $(n_s, r)$-plane with respect to the $\alpha$-attractor predictions.
The initial value of the axion $\theta_\ast$ determines the duration $N_{II}$ of the second stage of inflation.
The precise dependence can be computed by using the slow roll approximation of natural inflation:
$N_{II}(\theta_\ast) = -f^2_{a} \log\left(\frac{1-\cos(n\theta_\ast)}{2}\right).$
\end{itemize}
Some comments are in order. When the field reaches the valley of the Higgs potential, we may end inflation for a short while. This is more likely to happen if $A$ is really small, because its value determines the Hubble friction, which may or may not prevent inflation to end. We have checked numerically using methods of \cite{Dias:2015rca} that it does not affect the initial value of $\theta$ when natural inflation starts. Moreover, the field will oscillate strongly for a short amount of time. For fine-tuned initial conditions this might happen exactly during the observable regime and it will leave features in the spectra of perturbations \cite{Zelnikov:1991nv,Polarski:1992dq}. In this case the predictions become model dependent, and we cannot make definite statements.

\section {Appendix B: Other ways to generate natural inflation potentials}

By using the methods described in this paper as well as in \cite{Achucarro:2017ing,Yamada:2018nsk}  one can construct a very broad class of potentials, including the potentials of natural inflation with a weakly broken $U(1)$ symmetry, as well as the models where this symmetry is completely absent. Indeed, any potential $V(Z \bar Z)$ or $V(T+\bar T)$ is $U(1)$-symmetric, but our approach is equally valid for general potentials $V(Z,\bar Z)$ or $V(T,\bar T)$ without any $U(1)$ symmetry. Therefore one could construct the potentials of natural inflation with weakly broken $U(1)$ symmetry without modifying any parts of our construction but the potential $V$.

Other methods can also be used to reach a similar goal.  For example, one may start with a $U(1)$-symmetric potential $V(Z \bar Z)$ or $V(T+\bar T)$, and then add a small symmetry breaking term to the superpotential $W$. 

In particular, one can keep the $U(1)$ invariant function $V = V_0 \left(1-  c^{-2}Z\bar Z\right)^{2} +C$ of the T-model, and generate  the natural inflation potential by adding a small correction term $A\, Z$ to the superpotential $W = W_{0}$. Here $C$ is an arbitrary constant that can be added to the potential without affecting its $U(1)$ symmetry.  Similarly, one may consider the $U(1)$ invariant potential  $V = V_{0}\Big(1 - {T + \bar T\over 2}\, e^{\sqrt{2\over 3 \alpha}c}\Big)^{2} +C$ of the E-model, and generate the natural inflation potential by adding a small term $A\, e^{{-T}}$ to the superpotential.

We found that this method works well for $\alpha \leq 1/3$. However, if one considers $\alpha > 1/3$, the potential at large $\vp$ becomes unbounded from below. Thus one may either use the methods developed in the main part of our paper, or try to improve these models using singular potentials discussed in Section \ref{singul} and in \cite{Yamada:2018nsk}, or consider the models with $\alpha \leq 1/3$. Note  that the models with $\alpha = 1/3$ studied in \cite{Achucarro:2017ing} have a very interesting interpretation in terms of M-theory and extended supergravity and provide specific targets for the search of B-modes, as emphasized in \cite{Ferrara:2016fwe,Kallosh:2017ced}.

\newpage

\bibliographystyle{JHEP}
\bibliography{bibfile}

\end{document}